\documentclass[twocolumn,pra,showpacs]{revtex4-1}

\usepackage{amsmath}
\usepackage{graphicx}
\usepackage{color}

\usepackage{verbatim}

\newcommand{\nup}{N_{\uparrow}}
\newcommand{\up}{\uparrow}
\newcommand{\dn}{\downarrow}

\newcommand{\ket}[1]{\ensuremath{\left|#1\right\rangle}}

\begin{document}

\title{Fine structures in the spectrum of the open-boundary Heisenberg chain at large
  anisotropies}

\author{Auditya Sharma}
\affiliation{International Institute of Physics - Federal University of Rio Grande do Norte, Natal, RN, Brazil}
\author{Masudul Haque}
\affiliation{Max-Planck Institute for the Physics of Complex Systems, N\"{o}thnitzer Str. 38, 01187 Dresden, Germany}

\begin{abstract}

At large anisotropies, the spectrum of the Heisenberg XXZ spin chain separates into `bands' with
energies largely determined by the number of domain walls.  The band structure is richer with open
boundary conditions: there are more bands and the bands develop intricate fine structures.  We
characterize and explain these structures and substructures in the open-boundary chain.  The fine
structures are explained using degenerate perturbation theory.  We also present some dynamical
consequences of these sub-band structures, through explicit time evolution of the wavefunction from initial
states motivated by the fine structure analysis.

\end{abstract}

\maketitle

\section{Introduction}

Traditionally, the theory of many-body quantum systems has focused on the ground state and
low-energy parts of the eigenspectrum.  This is well-justified in solid-state systems which are
usually in contact with a thermal bath and typically relax fast to low-energy sectors.  As a
result, parts of the many-body eigenspectra away from the low-energy sector were generally
considered to be of no interest, for much of the history of condensed matter physics.

In recent years, the perspective has changed due to the advent of new experimental setups,
particularly those employing cold atom gases \cite{Cold-atom_reviews_expts}, which have promoted the
study of non-equilibrium situations in \emph{isolated} quantum systems \cite{noneq_reviews}.  In an
isolated situation, energy conservation ensures that a system with an initially high energy will not
reach the low-energy parts of the spectrum; the low-energy sector may thus be unimportant.  This
provides topical motivation to understand aspects of the full spectra of many-body systems.  In
particular, spectral structures in previously less-explored spectral regions can give rise to
unexpected dynamical phenomena \cite{haque2010self, JakschClark_PRB2013, KhomerikiFlach_various,
  Winkler-etal_Nature06, bound_clusters_itinerant, BHladder_Bloch_expt_theory,
  BoseHubbardDimer_various, MazzaFabrizio_PRB12, ChancellorPetriHaas_PRB13, superTG}.

In this work, we report and explain fine structures present in the eigenspectrum of the anisotropic
Heisenberg (XXZ) chain with open boundary conditions.  These spectral features are related to the
binding of magnons and the spatial relationship of bound multi-magnons with the edges of the open
chain.  Some related spectral fine structures were studied briefly in Ref.~\cite{haque2010self},
where it was also found that these structures give rise to unexpected dynamics suppression phenomena
(`edge-locking' of bound multi-magnons).  The study of this class of dynamics is now particularly
important because of the remarkable recent success in constructing spin chains in cold-atom setups
\cite{Fukuhara-2013a, Fukuhara-2013b, tilt_Isingspinchain}, which promises near-future experimental
explorations of many types of time evolution experiments with finite-size spin chains.

The XXZ chain is a fundamental model of condensed matter physics, and has long been the subject of
sustained theoretical activity.  The open chain has received less detailed attention than the
periodic case.  The Hamiltonian for an  $L$-site chain is 
\begin{equation}  \label{eq_Hamilt}
H_{XXZ} = J_x\sum_{j}\left[S_{j}^{x}S_{j+1}^{x}+S_{j}^{y}S_{j+1}^{y}+\Delta S_{j}^{z}S_{j+1}^{z}\right].
\end{equation}
The summation runs from $j=1$ to $j=L-1$ with open boundary conditions, and runs to $j=L$ with
periodic boundary conditions, the site $L+1$ being identified with site $j=1$.

The $S^zS^z$ term acts as an `interaction' penalizing alignment of neighboring spins.  The in-plane
terms $(S_j^xS_{j+1}^x+S_j^yS_{j+1}^y) = \tfrac{1}{2}(S_j^+S_{j+1}^- + S_j^-S_{j+1}^+)$ provide
`hopping' processes.  Since $H_{XXZ}$ preserves total $S^z$, the dynamics is always confined to
sectors of fixed numbers $N_{\uparrow}$ of up-spins.  For simplicity, our description will sometimes
focus on small $N_{\uparrow}$, i.e., highly polarized spin chains.  However, much of the phenomena
described here is aso valid at smaller magnetization (larger $N_{\uparrow}$).
We will mostly consider the large $\Delta$ regime, where the spectral structures are most prominent.
When not specified, energy and time are measured in units of $J_x$ and $\hbar/J_x$ respectively.

Refs.~\cite{Fukuhara-2013a, Fukuhara-2013b} have experimentally realized the XXZ chain Hamiltonian
\eqref{eq_Hamilt} with $\Delta\approx{1}$ using two hyperfine states of a bosonic species in a Mott
phase.  Ref.~\cite{Fukuhara-2013b} has also explored the propagation of bound multi-magnons, of
particular relevance to the physics described in the present manuscript.  A setup suitable for
realizing large $\Delta$ values is currently under development \cite{Gross_Bloch_private}.  
In addition, the XXZ model has been shown to describe Josephson junction arrays of the flux qubit
type \cite{LyakhovBruder_NJP05}, and may also be realizable in optical lattices
\cite{DuanDemlerLukin_PRL03} or with polaritons in coupled arrays of cavities
\cite{cca_KayAngelakis}.  It should be possible to explore the spectral structures and associated
dynamical phenomena described in the present article in one of these settings in the foreseeable
future.

At large $\Delta$, the number of domain walls (bonds connecting oppositely pointing spins) is a good
indicator of energy, and the spectrum accordingly splits up into energetically separated groups of
eigenstates or `bands'.  Section \ref{sec_overall_bands} describes this gross `band' structure and
the difference between open and periodic chains.  As pointed out in Ref.~\cite{haque2010self}, the
open chain spectrum not only has more bands than the periodic chain spectrum, but each of the bands
also has intricate sub-structures.  Ref.~\cite{haque2010self} studied the top two bands and
dynamical phenomena associated with them.  This is briefly reviewed in Section \ref{sec_top_two}.
The main object of this paper is to describe and explain the substructures in the bands lower
in the spectrum.  In Sections \ref{sec_third} and \ref{sec_fourth}, we explain the main sub-band
structures within the third and fourth bands from the top.  Section \ref{sec_dynamics_3rdband}
presents real-time evolution results which highlight the structure of the third band.

\section{\label{sec_overall_bands}  `Band' structure in many-body spectrum}

\begin{figure}[ht]
\includegraphics[width=0.99\columnwidth]{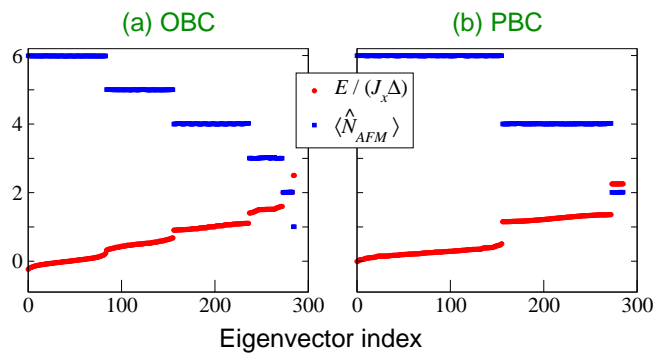}
\caption{Energy and $\langle \hat{N}_{AFM}\rangle$ (average number of domain walls) of eigenstates,
  indexed in order of increasing energy, for (a) open boundary conditions, and (b) periodic boundary
  conditions.  We show the $N_{\uparrow}=3$ sector, for a $L=13$ chain at large anisotropy
  $\Delta = 10$.  There are $2N_{\uparrow}=6$ bands in the OBC spectrum and $N_{\uparrow}=3$ bands
  in the PBC spectrum. }
\label{fig1}
\end{figure}

\begin{figure}
\includegraphics[width=0.99\columnwidth]{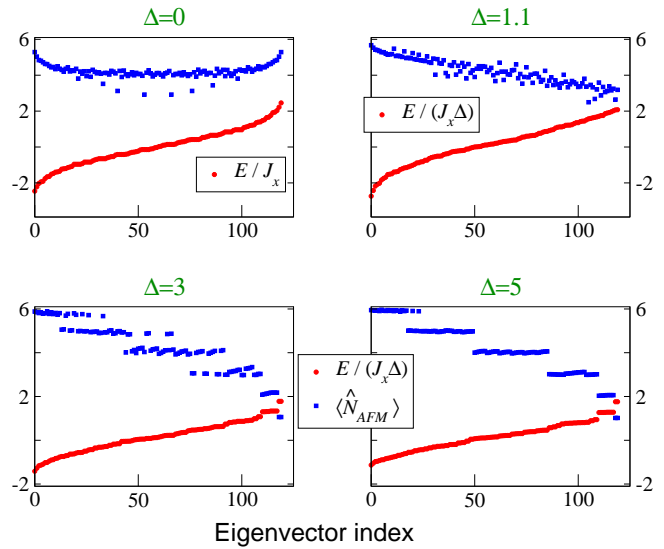}
\caption{\label{fig2} Energy and $\langle \hat{N}_{AFM}\rangle$ of eigenstates for $N_{\uparrow}=3$,
  $L=10$ chain with open boundary conditions for several $\Delta$ values.  The bands in energy
  start appearing at high $\Delta$; at moderate $\Delta$ the band structure is easier to see through
  the jumps in $\langle \hat{N}_{AFM}\rangle$.  }
\end{figure}

At $\Delta=\infty$, the spectrum splits into bands.  In this limit, the in-plane terms in the
Hamiltonian can be neglected, and we see a grouping of states into blocks that have different
numbers of `domain walls' or `anti-ferromagnetic nearest neighbors', $N_{AFM}$. 

The quantity $N_{AFM}$ can be defined as an operator
\begin{align}
\hat{N}_{AFM} ~=~ 
\sum_{j}(\frac{1}{2}-2 S_{j}^{z}S_{j+1}^{z}),
\end{align}
which is diagonal in the $S_z$ basis.  In the Ising limit this quantity has integer values for each
eigenstate.  The value depends on the boundary conditions (since the range of the summation index
$j$ does); e.g., the state $|\uparrow\uparrow\uparrow\downarrow\downarrow\cdots\downarrow\rangle$
has $N_{AFM} = 2$ for periodic boundary conditions, but it is only $N_{AFM}=1$ for open boundary
conditions.

For $N_{\uparrow}=3$, there are three bands in the periodic chain, corresponding to configurations
with (\emph{i}) the three $\uparrow$ spins next to each other ($N_{AFM}=2$), (\emph{ii}) two
$\uparrow$-spins neighboring each other and one disjoint ($N_{AFM}=4$), (\emph{iii}) the three
$\uparrow$ spins all non-neighboring ($N_{AFM}=6$).  In the open chain, the edge allows also
configurations with odd $N_{AFM}$ values; hence there are six bands.  The total number of states
in the $N_{\uparrow}=3$ sector is ${L\choose 3} =
\frac{1}{6}L(L-1)(L-2)$. Table~\ref{table1} collects the number of states in different bands for
both open and closed chains, and lists corresponding $N_{AFM}$ values.  These counting arguments can
be readily generalized to larger $N_{\uparrow}$ sectors.  

The band structure survives at finite but large anisotropies $\Delta$.  Figure \ref{fig1}
illustrates this for the $N_{\uparrow}=3$ sector with $\Delta=10$.  Also shown here is the
expectation value of $\hat{N}_{AFM}$ as defined above.  Althought the finite-$\Delta$ eigenstates of
the Hamiltonian are not eigenstates of the operator $\hat{N}_{AFM}$, the expectation value of
$\hat{N}_{AFM}$ nevertheless remains close to the integer value expected for $\Delta \to \infty$,
and there are sharp jumps of $\langle\hat{N}_{AFM}\rangle$ across bands.

To show the extent to which the band structure survives at smaller $\Delta$, Figure \ref{fig2}
collects data for the open chain for a range of values of $\Delta$.  We see that $\langle
\hat{N}_{AFM}\rangle$ is a rather sharp indicator of the band structure.

\begin{table}
  \begin{tabular}{|l|l|l|}
    \hline
    $N_{AFM}$ & Number of states(OBC) & Number of states(PBC)\\
    \hline
    1  & 2 & -\\
    2  & L-2 &  L\\
    3  & 4(L-4) & -\\
    4  & (L-4)(L-4) & L(L-4)\\
    5  & (L-4)(L-5) & -\\
    6  & (L-4)(L-5)(L-6)/6 & L(L-4)(L-5)/6\\
    \hline
\end{tabular}
\caption{The number of states belonging to different `bands' at large $\Delta$ within the $\nup=3$
  sector.  Each band being labeled by a value of $N_{AFM}$.  The numbers sum to the total number of
  states in the full $N_{\uparrow}=3$ sector, $\frac{L(L-1)(L-2)}{6}$. There are only three
  bands for the closed chain, corresponding to $N_{AFM}=2$,$4$, and $6$, and six bands for the open chain.}
\label{table1}
\end{table}

\subsection{\label{sec_top_two} Top two bands}

The top two bands, corresponding to $\langle\hat{N}_{AFM}\rangle\approx1$ and to
$\langle\hat{N}_{AFM}\rangle\approx2$, were considered in Ref.\ \cite{haque2010self}.  Both have the
same number of states for arbitrary $N_{\uparrow}$, namely $2$, and $L-2$.  The application of the
hopping ($J_x$) terms does not lift the degeneracy of these two bands within first
order-perturbation theory; thus the substructures of the $\langle\hat{N}_{AFM}\rangle\approx2$ only
appear at higher order.  We briefly review the physics of these bands here.

The top band is made of configurations where all $\nup$ up-spins are at one edge.  There are two
states in this band because the up-spins can be pinned either to the left or the right edge.  The
separation of this band from the rest of the spectrum leads to the ``trivial edge-locking'' physics
described in Ref.\ \cite{haque2010self}: a configuration like $\ket{\up\up\dn\dn\dn...}$  will have
very litte dynamics at large $\Delta$ as the $\up$-block is locked to the edge.  

The second band is composed of (a) $\up$-spins in a single block away from the boundaries, (b) the
$\up$-spins separated into two blocks, each of them pinned to one of the edges of the chain.  The
band has substructures appearing at second and higher even orders in the hopping term; these lead to
a `fractal' structure in the specturm and a corresponding set of nontrivial edge-locking phenomena
\cite{haque2010self}.

\begin{figure}
\centering \includegraphics[width=0.99\columnwidth]{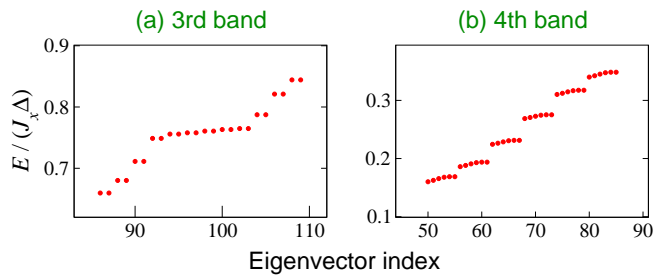}
\caption{ Blow-ups of third and fourth band from the top, for an open $L=10$ chain in the
  $N_{\uparrow}=3$ sector; $\Delta=10$.  (a) Third band, containing $4(L-4) = 24$ states, consists of a flat
  patch between two staircases of nearly degenerate pairs.  (b) Fourth band has $(L-4)$ steps each
  containing  $(L-4)$ states, thus having a total of  $(L-4)^2 = 36$ states.    }
\label{fig3}
\end{figure}

\subsection{This work: substructures in third and fourth bands from the top}

In this work, we will focus on the substructures visible in the next two (third and fourth) bands
from the top.  In both cases, substrucutres already appear at first order in the hopping term.  We
introduce briefly the first-order substructures here, and analyze them in detail in the remainder of
the article.

The third band contains $(L-3)$ nearly degenerate paris of states.  For $\nup=2$, this exhausts the
band.  For any $\nup>2$, there is a nearly degenerate central group which contains the rest of the
states of the third band.  Figure \ref{fig3}(a) shows an example blow-up of the third band, for
$\nup=3$ and $L=10$.

For any $\nup>2$, the fourth band contains $(L-4)$ almost degenerate groups of $(L-4)$ states.  For
$\nup=3$, this exhausts the band, as seen in Figure \ref{fig3}(b).  For any $\nup>3$, there are more
than $(L-4)^2$ states in the fourth band, so there is a nearly degenerate central group which
contains the remainder.  This will be seen in Figure \ref{fig7}.
(The $\nup=2$ case is special for the fourth band and is discussed briefly at the end of Section
\ref{sec_fourth}.)

\section{\label{sec_third} Third band from the top}

In this section we will consider and explain the substructures seen in the third band from the top,
corresponding to $\langle\hat{N}_{AFM}\rangle\approx3$ at large $\Delta$, for arbitrary
$N_{\uparrow}\le L/2$.  We can only get configurations with $N_{AFM}=3$ if the $N_{\uparrow}$
up-spins appear in exactly two isolated blocks, and exactly one of these blocks is pinned to one of
the edges.  We will refer to the two blocks as the `edge-pinned' and `free' blocks.  With these
constraints, we can count that the number of states in the third band is
$2(N_{\uparrow}-1)(L-N_{\uparrow}-1)$. For $N_{\uparrow}=3$, this is $4(L-4)$, which agrees with
Table.~\ref{table1}.

\begin{figure}
\centering \includegraphics[width=0.99\columnwidth]{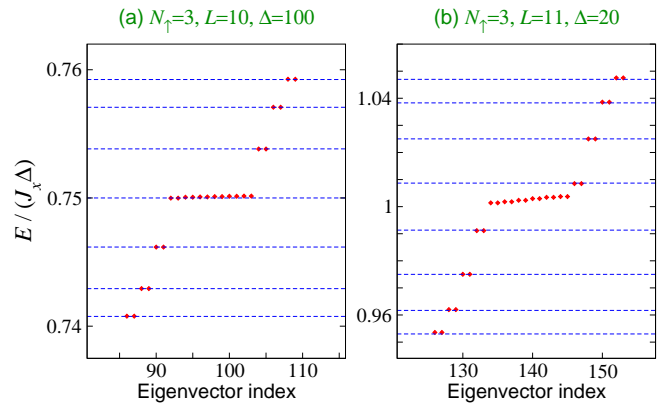}
\caption{\label{fig4}
Third band from the top, $N_{\uparrow}=3$ sector, for open chains
of size $L=10$ and $L=11$.  Dashed horizontal lines are from first-order degenerate perturbation
theory, $\epsilon_{3}+\lambda e_{r}$.  (a) Perturbation theory is seen to be quite accurate for
$\Delta = 100$.  (b) At smaller $\Delta$, there are visible deviations from perturbation theory and
a dispersion in the central `flat' patch.  For the even case $L=10$, one of the $e_{r}$'s is zero,
so that the central patch has length $2(L-4)=12$.  For the odd case $L=11$, the central group
contains $2(L-5)=12$ states.
 }
\end{figure}

Now, let us look at the substructure of this band.  Figure \ref{fig3}(a) blows up the region around
the third band for open chains of size $L=10$ for the $N_{\uparrow}=3$ sector. We see a central flat
patch where a group of eigenstates are placed at roughly the same energies. On either side of this
flat region, we find stairs with nearly degenerate pairs of eigenstates.  We will next show that the
substructure of this band for arbitrary $N_{\uparrow}$ can be understood using first-order
degenerate perturbation theory.

We start by rewriting the original Hamiltonian as
\begin{align}
\frac{H}{J_{x}\Delta} = H_{0}+\lambda H^{'},
\end{align}
where $H_{0}=\sum_{j}S_{j}^{z}S_{j+1}^{z}$,
$H^{'}=\sum_{j}S_{j}^{x}S_{j+1}^{x}+S_{j}^{y}S_{j+1}^{y}$, and $\lambda = 1/\Delta$. We are
interested in the third band from the top characterized by states with $N_{AFM}=3$ being dominant.
In the Ising limit ($\lambda=0$), all the states in this band have identical energy $\epsilon_{3} =
(L-7)/4$ (in units of $J_{x}\Delta$), because $H_{0}$ depends only on $N_{AFM}$. 
For example, the third band appears around $0.75J_x\Delta$ for $L=10$ and around  $J_x\Delta$ for
$L=11$ (Figures \ref{fig3}(a) and \ref{fig4}).

Now, we turn on a tiny $\lambda > 0$.  Exploiting standard degenerate perturbation theory, we recall
that we must diagonalize the matrix $W_{ij} = \langle\psi_{i}^{0}|H'|\psi_{j}^{0}\rangle$, where
$|\psi_{i}^{0}\rangle$ are the $2(N_{\uparrow}-1)(L-N_{\uparrow}-1)$ degenerate eigenstates that
constitute the third band when $\lambda = 0$.  This matrix nicely separates into two identical
blocks, corresponding to whether the `pinned' group of $\uparrow$ spins is at the right edge or at
the left edge.  Being identical, these blocks give the same eigenvalues, and thus accounts for the
occurrence of degenerate pairs in Fig.\ref{fig3}. We will now concentrate on the eigenvalues of the
block which has the left end `pinned'.

Among the `left-pinned' states, only $(L-3)$ states have first-order matrix elelments.  This is
independent of $N_{\uparrow}$, so that the nonzero part of the $W_{\rm left}$ matrix is the same
regardless of $N_{\uparrow}$.  The remaining $(N_{\uparrow}-2)(L-N_{\uparrow}-2)$ states give a
large block of zeros at first order, and are responsible for the flat patch at the middle of the
third band.    

The states connected at first order by $H'$ include the $L-\nup-1$ configurations where $\nup-1$
up-spins are edge-pinned and one is free, e.g., for $\nup=4$, the states 
$|\uparrow\uparrow\uparrow\downarrow\downarrow\cdots\downarrow\uparrow\downarrow\rangle$,
$|\uparrow\uparrow\uparrow\downarrow\downarrow\cdots\downarrow\uparrow\downarrow\downarrow\rangle$,$|\uparrow\uparrow\uparrow\downarrow\downarrow\cdots\downarrow\uparrow\downarrow\downarrow\downarrow\rangle$,
$\cdots$, $|\uparrow\uparrow\uparrow\downarrow\uparrow\downarrow\cdots\downarrow\rangle$.  Any two
successive pairs of these are connected by $H'$.  Successive application of $H'$ on the last of
these states connects also configurations with smaller numbers $\nup-r$ of edge-pinned spins (one
for each $r>1$).  In the $\nup=4$ case, these are 
$|\uparrow\uparrow\downarrow\uparrow\uparrow\downarrow\cdots\downarrow\rangle$ and 
$|\uparrow\downarrow\uparrow\uparrow\uparrow\downarrow\cdots\downarrow\rangle$.   There are $\nup-2$
such states.  So the total number of states that give a non-zero block in
$W_{\rm left}$ is $L-N_{\uparrow}-1+N_{\uparrow}-2 = L-3$, independent of $N_{\uparrow}$. We write
the $(L-3)\times(L-3)$ dimensional matrix $W_{\rm left}$:
\begin{align}
W_{\text{left}} ~=~ T_{(L-1)} ~=~ 
\begin{pmatrix}
0 & \frac{1}{2} & 0 & \cdots & 0 & 0 & 0\\
\frac{1}{2} & 0 & \frac{1}{2} & 0 & \cdots & 0 & 0\\
0 & \frac{1}{2} & 0 & \frac{1}{2} & 0 & \cdots & 0\\
\vdots\\
0 & 0 & \cdots & 0 & \frac{1}{2} & 0 & \frac{1}{2}\\
0 & 0 & \cdots & 0 & 0 & \frac{1}{2} & 0
\end{pmatrix} .
\label{eq:W_left_generic}
\end{align}
This is a well-studied type of matrix called a $1$-Toeplitz matrix \cite{gover1994eigenproblem}; we
will denote a $m\times{m}$ $1$-Toeplitz matrix as $T_m$.  Its eigenvalues are
\begin{align} \label{eq_3rdband_cosine}
e_{r} = -\cos(\frac{r\pi}{L-2}), \;\;\;\;\;\;\; r=1,2,\cdots,(L-3).
\end{align}
The number of steps in the third band and their positions $\epsilon_3+\lambda{e}_{r}$ are thus
fixed by $L$ and $\Delta$, independent of $N_\uparrow$.  

When $L$ is even, a zero eigenvalue appears because of $\cos{\frac{\pi}{2}}$. The number of states
appearing at $\epsilon_3$ is thus $2((N_{\uparrow}-2)(L-N_{\uparrow}-2)+1)$ for even $L$ and
$2(N_{\uparrow}-2)(L-N_{\uparrow}-2)$ for odd $L$.

In Figure \ref{fig4} we compare third-band substructures at $\Delta=100$ and $\Delta=20$ with
degenerate perturbation theory, for the $\nup=3$ sector.  For $\nup=3$, the number of states in the
central flat patch (``zeros block'') is $2(L-4)$ for even $L$  and  $2(L-5)$ for odd $L$.

\section{\label{sec_dynamics_3rdband} Propagation Dynamics}

\begin{figure*}
\centering \includegraphics[width=0.8\textwidth]{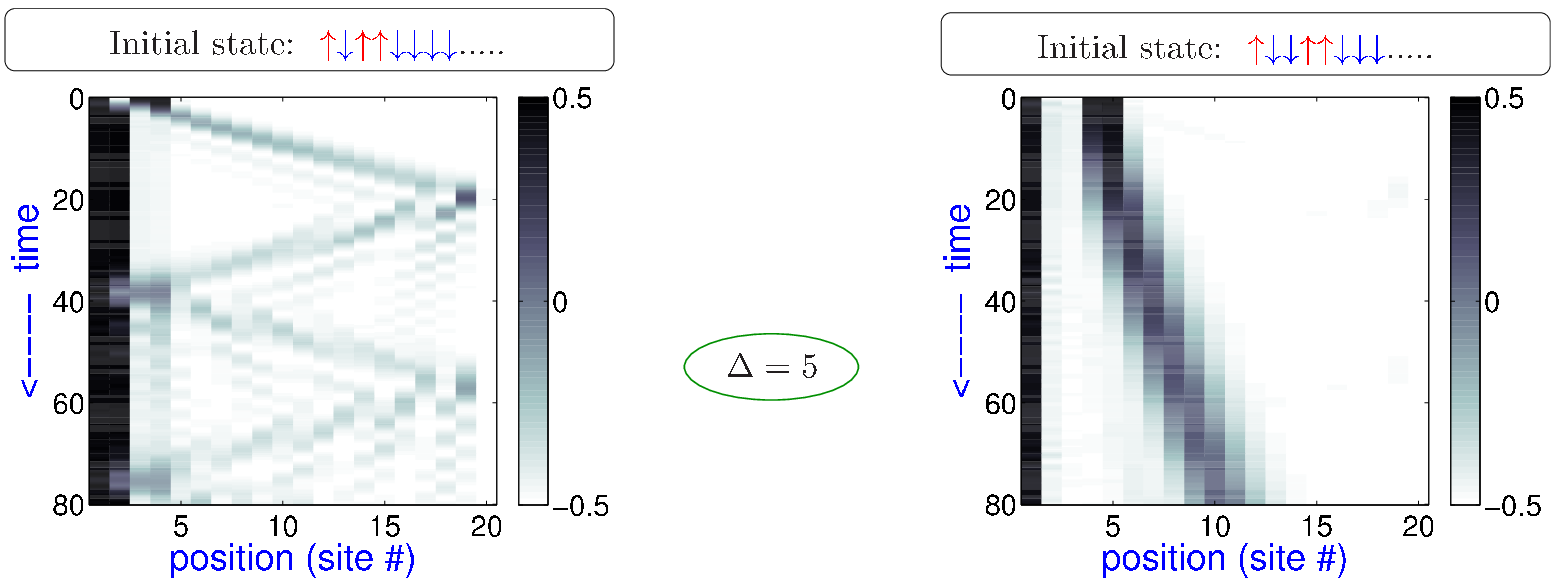}
\caption{ \label{fig_dynamics_3rdband} Time evolution for chains initiated with configurations
  $\ket{\alpha}$ and $\ket{\beta}$, as defined in Equation \eqref{eq_init_configs}.  Time units are
  $J_x^{-1}$; $\Delta=5$.  For initial state $\ket{\alpha}=\ket{\up\dn\up\up\dn\dn\dn...}$, the
  propagation is fast as a single magnon.  The initial state
  $\ket{\beta}=\ket{\up\dn\dn\up\up\dn\dn...}$ is part of the `flat' patch of the third band, not
  connected by first-order hopping processes; dynamics therefore involves slower bi-magnon
  propagation.
}
\end{figure*}

Having analyzed the structure of the states making up the third band from the top (band with
$\langle\hat{N}_{AFM}\rangle\approx3$), we can explore some dynamical consequences (Figure
\ref{fig_dynamics_3rdband}).  Considering the $\nup=3$ case, we have found above that all but one of
the configurations with one $\uparrow$ spin edge-pinned are part of the flat manifold due to having
no matrix elements at leading order.  The exception is the state $\ket{\up\dn\up\up\dn\dn...}$,
i.e., the configuration where the `free' group of $\nup-1$ spins sits at minimum distance from the
edge-pinned part.  This configuration is part of the nonzero $W$ matrix constructed in the last
section, and hybridizes at first order with the configurations where a single $\up$ spin is free.
This implies, for example, that open chains initiated with states
\begin{equation} \label{eq_init_configs}
\ket{\alpha}=\ket{\up\dn\up\up\dn\dn\dn...} \quad \mathrm{and} \quad \ket{\beta}=\ket{\up\dn\dn\up\up\dn\dn...}
\end{equation}
should have qualitatively different dynamics: the first can propagate by first-order hopping
processes and the second cannot.

Figure \ref{fig_dynamics_3rdband} displays this difference.  Starting with the initial state
$\ket{\alpha}$, the `free' block transfers ($\nup-2=1$) of its members to the edge-pinned block,
after which the remaining single free $\up$ spin propagates at the typical magnon velocity, $J_x$
\cite{GanahlEsslerEvertz_PRL12}.  With the initial state $\ket{\beta}$, however, the dynamics does
not significantly access the ``single-free-spin'' sector; therefore the propagation is through
higher-order processes; the non-edge-pinned part moves as a bound bi-magnon at slower speeds
\cite{GanahlEsslerEvertz_PRL12}.

\section{\label{sec_fourth} Fourth Band from the top}

\begin{table}
  \begin{tabular}{|c|c|}
    \hline
    type (a) & type (b) \\
    \hline
    $\quad\up\up\dn\dn\dn\dn\dn\dn\dn\up\dn\up\quad$  &     $\quad\dn\dn\dn\up\up\up\dn\dn\dn\dn\up\dn\quad$  \\
    $\quad\up\up\dn\dn\dn\dn\dn\dn\up\dn\dn\up\quad$  &     $\quad\dn\dn\dn\up\up\up\dn\dn\dn\up\dn\dn\quad$ \\
    $\quad\up\up\dn\dn\dn\dn\dn\up\dn\dn\dn\up\quad$  &     $\quad\dn\dn\dn\up\up\up\dn\dn\up\dn\dn\dn\quad$ \\
    $\quad\up\up\dn\dn\dn\dn\up\dn\dn\dn\dn\up\quad$  &     $\quad\dn\dn\dn\up\up\up\dn\up\dn\dn\dn\dn\quad$ \\
    $\quad\up\up\dn\dn\dn\up\dn\dn\dn\dn\dn\up\quad$  &     $\quad\dn\dn\dn\up\up\dn\up\up\dn\dn\dn\dn\quad$ \\
    $\quad\up\up\dn\dn\up\dn\dn\dn\dn\dn\dn\up\quad$  &     $\quad\dn\dn\dn\up\dn\up\up\up\dn\dn\dn\dn\quad$ \\
    $\quad\up\up\dn\up\dn\dn\dn\dn\dn\dn\dn\up\quad$  &     $\quad\dn\dn\up\dn\dn\up\up\up\dn\dn\dn\dn\quad$ \\
    $\quad\up\dn\up\up\dn\dn\dn\dn\dn\dn\dn\up\quad$  &     $\quad\dn\up\dn\dn\dn\up\up\up\dn\dn\dn\dn\quad$ \\
    \hline
\end{tabular}
\caption{\label{table2} Examples of groups of $(L-4)$ states in the fourth band which are connected by
  first-order hopping processes and hence form 1-Toeplitz blocks $T_{(L-4)}$ in the degeneracy matrix $W$.  Left
  [right] column shows a group of (a) type [(b) type] configurations.  These examples are shown for
  $\nup=4$, $L=12$.
}
\end{table}

Next we will consider the fourth band from the top, corresponding to
$\langle\hat{N}_{AFM}\rangle\approx 4$ at large $\Delta$.  This band does not appear for
$N_{\uparrow}=1$ and is the lowest (last) band for $N_{\uparrow}=2$, discussed briefly at the end of
the section.  

For $3\le N_{\uparrow}\le L/2$, we can only get configurations with $N_{AFM}=4$ in one of the
following two ways:
\begin{itemize}
\item[(a)] There are three isolated groups of contiguous $\up$-spins, one pinned to the left end of
  the chain, one to the right end, and the third in between not pinned to either end.
\item[(b)] There are two isolated groups of contiguous $\up$-spins, neither of which is pinned to an
  end. 
\end{itemize}
To count the number of configurations of type (a), let us say there are $r$ up-spins on the left
end, and $s$ up-spins on the right end (with $r\ge 1$,$s\ge 1$), and $t$ up-spins in between, such
that $r+s+t=N_{\uparrow}$.  The number of ways of choosing $r$ and $s$ is
$\binom{N_{\uparrow}-1}{2}=\frac{1}{2}(N_{\uparrow}-1)(N_{\uparrow}-2)$. For each of these ways of
choosing $r,s$, there are $L-N_{\uparrow}-1$ ways of putting the central block of $t$-up-spins. Thus
there are $\frac{1}{2}(N_{\uparrow}-1)(N_{\uparrow}-2)(L-N_{\uparrow}-1)$ states of type (a).  In
order to count the number of ways of doing case (b), we first observe that there are
$(N_{\uparrow}-1)$ ways in which $N_{\uparrow}$ can be divided between the two groups. For each of
these ways of division, there are $\frac{1}{2}(L-2-N_{\uparrow})(L-1-N_{\uparrow})$ ways of
positioning the two groups in the available $L-2$ spots. Thus the total number of states of type (b)
is $\frac{1}{2}(N_{\uparrow}-1)(L-2-N_{\uparrow})(L-1-N_{\uparrow})$. Therefore, the total number of
states in this band is $\frac{1}{2}(N_{\uparrow}-1)(L-N_{\uparrow}-1)(L-4)$.  For $\nup=3$, this
reduces to $(L-4)^2$ (Table \ref{table1}).

We will next examine and explain the leading (first-order) substructure of this band.  Figure
\ref{fig3}(b) zooms into the fourth band for $\nup=3$: we see $(L-4)$ subbands each containing
$(L-4)$ states.  The same subband structure persists at larger $\nup$; however for $\nup\geq4$ there
is also a ``flat patch'' at $\epsilon_{4} = (L-9)/4$, analogous to the flat patch encountered above
for the third band.

\begin{figure}
\centering \includegraphics[width=0.99\columnwidth]{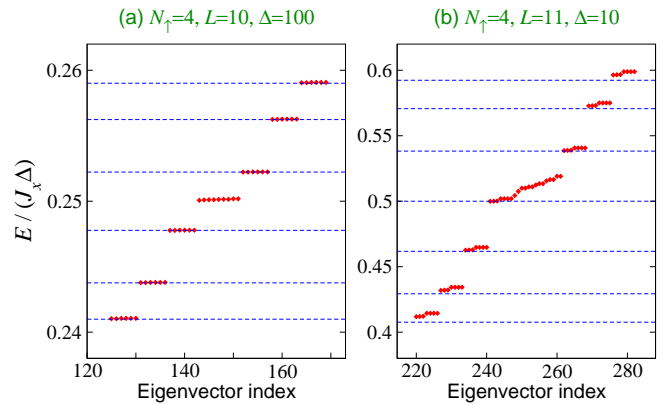}
\caption{\label{fig7}
Fourth band from the top of the $N_{\uparrow}=4$ spectrum, for open chains of size $L=10$ and
$L=11$.  Dashed horizontal lines are from first-order degenerate perturbation theory,
$\epsilon_{4}+\lambda e_{r}$.  Panel (a) has a large anisotropy, $\Delta = 100$, so that the
perturbative calculations are quite accurate.  At smaller $\Delta = 10$, the essential structure is
the same, but each subband now deviates slightly from the first-order prediction, has visible
dispersion, and also visible higher-order sub-structures.  One of the groups of $(L-4)$ merges with
the ``zeros block'' at $\epsilon_{4}=(L-9)/4$ for the odd-sized chain (b), but not for the even-site
chain (a).  
}
\end{figure}

As done for the third band, we use first-order degenerate perturbation theory.  Again, we start by
writing the original Hamiltonian as $H/(J_{x}\Delta) = H_{0}+\lambda H^{'}$, where
$H_{0}=\sum_{j}S_{j}^{z}S_{j+1}^{z}$, $H^{'}=\sum_{j}S_{j}^{x}S_{j+1}^{x}+S_{j}^{y}S_{j+1}^{y}$, and
$\lambda = 1/\Delta$.  When $\lambda=0$, all the states in the $\langle\hat{N}_{AFM}\rangle= 4$ band
have identical energy $\epsilon_{4} = (L-9)/4$ in units of $J_{x}\Delta$, regardless of
$N_{\uparrow}$.  
(E.g., In Figures \ref{fig3}(b) and \ref{fig7}(a) the fourth band is seen to be around
$E\sim0.25J_x\Delta$ for $L=10$.)
Again, we have the task of diagonalizing the
matrix $W_{ij} = \langle\psi_{i}^{0}|H|\psi_{j}^{0}\rangle$, where $|\psi_{i}^{0}\rangle$ are the
$\frac{1}{2}(N_{\uparrow}-1)(L-N_{\uparrow}-1)(L-4)$ degenerate eigenstates that constitute the
fourth band when $\lambda = 0$.  We will show below that the matrix $W$ can entirely be written as
blocks of 1-Toeplitz matrices $T_{m}$, and that there are $(L-4)$ such blocks each having size
$(L-4)$, independent of $N_{\uparrow}\geq3$.  Of the $(L-4)$ blocks, $(N_{\uparrow}-2)$ are made 
of configurations of type (a) and $L-\nup-2$ are composed of  configurations of type (b).

Configurations of type (a) and type (b) do not mix at first order and naturally separate into
blocks. Table \ref{table2} (left) shows an example of $(L-4)$ type-(a) configurations which are
connected by single hoppings.  Most of these states have a single `free' $\up$ spin in the interior.
The number of ways of splitting the remaining $\nup-1$ spins into left and right blocks is $\nup-2$;
e.g., for $\nup=4$ we can have two $\up$'s at the left edge and one $\up$ at the right, or one $\up$
at the left edge and two at the right.  There are thus $\nup-2$ Toeplitz blocks $T_{(L-4)}$
contributed by (a)-type configurations to the $W$ matrix.

We now turn to the contributions from (b) type configurations.  Table \ref{table2} (right) shows an
example of $(L-4)$ type-(b) configurations connected by single hoppings, so that they form a
$T_{(L-4)}$ block in the $W$ matrix.  If thought of as a time evolution sequence, this is a
``quantum bowling'' event as studied in Ref.\ \cite{GanahlHaqueEvertz_arxiv13}, where a single
`particle' ($\up$-spin) passes through a `wall' of $\nup-1$ particles, by turning into a `hole'
while during the transmission and shifting the wall by two sites in the process.  All the
contributing states of  type (b) fall into this class of blocks.  The number of ways in which one
can arrange the $\nup$ $\up$-spins into a block of $(\nup-1)$ to the left of a single $\up$-spin on
the second-rightmost site, is  $(L-N_{\uparrow}-2)$; as a result there are  $(L-N_{\uparrow}-2)$
Toeplitz blocks from (b) type configurations. 

Therefore, we now see that the nonzero part of the $W$ matrix is composed of
$N_{\uparrow}-2+L-N_{\uparrow}-2=(L-4)$ blocks of $T_{(L-4)}$: 
\begin{align}
W =
\begin{pmatrix}
T_{(L-4)} & 0 & 0 & \cdots & 0 & 0 & 0\\
0 & T_{(L-4)} & 0 & 0 & \cdots & 0 & 0\\
0 & 0 & T_{(L-4)} & 0 & 0 & \cdots & 0\\
\vdots\\
0 & 0 & \cdots & 0 & 0 & T_{(L-4)} & 0\\
0 & 0 & \cdots & 0 & 0 & 0 & T_{(L-4)}
\end{pmatrix} .
\end{align}
(The zero blocks are omitted.)  Once again using the results of $1$-Toeplitz
matrices~\cite{gover1994eigenproblem}, the eigenvalues of $W$ are found to be
\begin{align}
\label{eqn:band4pert}
e_{r} = -\cos(\frac{r\pi}{L-3}), \;\;\;\;\;\;\; r=1,2,\cdots,(L-4),
\end{align}
with each eigenvalue being $(L-4)$ times degenerate.  So, the energies of the sub-bands of the fourth
band should be given by $\epsilon_{4}+\lambda e_{r}$.

The nonzero part  of the $W$-matrix has size $(L-4)^{2}$, which is less than the number of states,
$\frac{(N_{\uparrow}-1)(L-N_{\uparrow}-1)(L-4)}{2}$, in the fourth band, for any $\nup\geq4$.
Therefore, in addition to the $(L-4)$ groups of states at energies  $\epsilon_{4}+\lambda e_{r}$,
there is an additional group of states at energy  $\epsilon_{4}$.  For odd $L$,the middle of the
$(L-4)$ groups falls at  $\epsilon_{4}$ (because $e_{r=(L-3)/2}=0$) and adds to the central patch.  

In Figure \ref{fig7}, we show a comparison  for the $\nup=4$ sector between perturbative results
and the actual spectra.  Our explanation based on first-order perturbation theory captures the
essential features even for anistoropies as low as $\Delta\sim10$.

For $\nup=3$, there is no zeros block because the $(L-4)$ groups of $(L-4)$ states each exhaust the
fourth band; $\frac{(N_{\uparrow}-1)(L-N_{\uparrow}-1)(L-4)}{2} = (L-4)^{2}$.  An example spectrum appeared
in Figure \ref{fig3}(b).

\subsection{$N_{\uparrow}=2$}

For $N_{\uparrow}=2$, the fourth band from the top is the `last' (lowest-energy) band and does not
have the substructure of $L-4$ steps.  Examining the $W$ matrix shows why this case is different.
For $\nup=2$, only configurations of type (b) occur: these contain two isolated single $\up$ spins
which can both move by first-order processes.  Hence the $W$ matrix of first order couplings will
have four entries on most rows, so that it is not of 1-Toeplitz type.  The problem is equivalent to
two spinless fermions on an open $(L-2)$-site chain with forbidden nearest-neighbor occupancy.  The
spectrum is thus quite different from the fourth-band structures for $\nup\geq3$.

\section{\label{sec_farther} Bands farther from the top; higher-order structures}

The bands farther down in the spectrum have less pronounced structure.  For the
$\langle\hat{N}_{AFM}\rangle\approx5$ band, degenerate perturbation theory at first order leads to a
complicated degeneracy matrix which is similar to that encountered in the pathological
$N_{\uparrow}=2$ case of the fourth band; numerical diagonalization of this matrix reproduces the
actual spectrum at large $\Delta$, but we were unable to obtain simple analytic expressions.

We have focused on subband structures that appear at first order in the hopping ($J_x$) term.  As in
the second band from the top \cite{haque2010self}, there are various higher-order substructures
present in the third and fourth bands, especially for larger $\nup$.  A full exploration is beyond
the scope of this work.  However, we make a couple of higher-order observations here.  First, within
the $(L-4)$-groups of the 4-th band, in Figure \ref{fig7}(b) we can see a separating out of four
states from the rest, at order $\lambda^2$, just as happens in the second band.  In fact, just as in
the second band \cite{haque2010self}, for larger $L$ this is the beginning of a recursive splitting,
where 4 levels separate out from the rest of the $(L-4)$, $(L-8)$,... levels at order $\lambda^2$,
$\lambda^4$,... until the $(L-4)$ states are exhausted.  Second, within the central patch of the
third band which is a zeros block at first order, one can zoom in to find pairs of states forming
second order dispersions.  These correspond to the motion of bound bi-magnons, which are connected
by second-order hoppings.

\section{Conclusions}

We studied band structures and substructures in the open-boundary Heisenberg chain at large
anisotropy, using degenerate perturbation theory to explain the most prominent features.  We also
presented real-time evolution phenomena reflecting these spectral structures.  

The features of the spectrum we have examined relate in particular to the propagation and binding of
magnons and collections of magnons in a ferromagnetic background.  The study of this type of
non-equilibrium situations is of growing current interest \cite{haque2010self,
  GanahlEsslerEvertz_PRL12, GanahlHaqueEvertz_arxiv13, WoellertHonecker_PRB12, Fukuhara-2013b}.  The
intriguing possibility of controlling such propagation phenomena using geometric features like edges
and impurities deserves further exploration; our dynamics results in Section
\ref{sec_dynamics_3rdband} is just one such example.  Spectral structures such as band structures,
and edge-locking phenomena, have also been found to be relevant to intriguing transport phenomena
\cite{ProsenRossiniZnidaric_PRB09, JakschClark_PRB2013}.

Our work also raises several other questions, e.g., are there generic ways in which the spectral
structures get modified at smaller $\Delta$?  It is also an open question how the intricate spectral
structures get modified or destroyed if we have a weak bond rather than a fully open-boundary chain;
one can interpolate between periodic and open-boundary cases by weakening one of the bonds in a
periodic chain.  
We have presented a real-time dynamical effect in Section \ref{sec_dynamics_3rdband} related to the
first-order substructure of the third band.  There should be various classes of non-equilibirum
effects associated with the higher-order structures of both third and fourth bands, which may be
worth exploring in detail.
Finally, since the open XXZ chain is integrable via Bethe ansatz, intricate
structures in the spectrum are reflected in intricate structures in the Bethe ansatz roots, which
sometimes show imaginary-real transmutations as a function of $\Delta$, as has been worked out for
the $\nup=2$ sector in Ref.\ \cite{AlbaSahaHaque_BetheAnsatz}.  Examining the Bethe ansatz
description of edge-related states for higher $\nup$ sectors remains an open task.

\begin{acknowledgments}

AS thanks S.~Ramasesha for valuable discussions, and a careful reading of the manuscript.

\end{acknowledgments}

\end{document}